# Topological Analysis of Foams and tetrahedral structures

**GAD FRENKEL[1], RAFI BLUMENFELD[1,2], PETER R. KING[1], MARTIN BLUNT[1]**

[1]*Earth Science and Engineering, Imperial College, London SW7 2AZ, UK.*
[2]*Biological and Soft Systems, Cavendish Laboratory, Cambridge CB3 0HE, UK.*

## ABSTRACT

In this paper we characterize foams and tetrahedral structures in a unified way, by a simplified representation of both that conserves the system's topology. The paper presents a workflow for an automated characterization of the topology of the void space, using a partition of the void space into polyhedral cells connected by windows. This characterization serves as the basic input for the Edwards entropic formalism that deals with the statistical characterization of configurational disorder in granular aggregates and argued to work for foams. The Edwards formalism is introduced and simplified expectation-values are calculated.

## INTRODUCTION

Foams and granular media have mutual properties. Both have disordered structure, for both the void space is as important as the material and for both the characterization of the structure is a key to the understanding and prediction of physical properties, such as permeability and thermal or electrical conductivity. These similarities suggest that a unified formalism should be found to treat the structure and properties of these systems. Such formalism is found in a paper by Edwards and Oakeshott [1] that conjectured that granular materials can be treated via an entropic formalism, which is similar to statistical mechanics, provided that the energy of the system is replaced by its volume. The key to the application of the formalism is the description of the states of the system by a set of parameters called the degrees of freedom.

In this paper we present the first few steps in the unification of foams and granular aggregates. We start by the transformation of foam to a structure composed of pseudo-grains. Next an automated method is described to identify the cellular structure. The cellular structure is used to extract topological data such as the number of windows per cell. Last, we show for a simplified 2D structure how to use the Edwards formalism to obtain expectation values. However, to put everything in context we start with a short introduction to the Edwards Formalism.

## EDWARDS' COMPACTIVITY

The statistical mechanical formalism of Edwards is based on the replacement of the energy of thermodynamic systems by volume and of temperature by an analogue quantity – compactivity [1, 2]. In this approach the system is assumed to have a very large set of possible 'states', where each state represents a possible configuration of the system components. For grains a state is defined by the position of grains, orientation, shape etc. For foam a state is defined by the cellular structure, the connectivity of cells etc. These parameters are called the degrees of freedom because their distribution determines the possible states. Each state has a probability of occurrence that depends on the system's volume $V$ and states with identical volume are assumed to have equal probabilities of occurrence. The probability factor is given by the modified Boltzmann factor $\exp(-v/x)$ where the parameter $X$ was named as the Compactivity. Since the phase space of the system's states is continuous, the configuration of parameters is accompanied by a density of states $\Theta(\{q\})$, which counts the number of states in the differential volume around the configuration of parameters $\{q\}$. The density of states also throws non-physical and non-available states.

The partition function is an non-normalized integral over the probabilities of all possible states of the system, which are expressed by the set of parameters $\{q\}$,

$$Z = \int e^{-W(\{q\})/x} \Theta(\{q\}) \Pi dq \,. \qquad (1)$$

In eq. (1), $W$ is the volume function that gives the volume of the entire system in terms of all the independent parameters in the system $\{q\}$. The entropy is defined as the log of the number of states at given volume. In terms of the entropy, the compactivity is defined as $X = \partial W / \partial S$; in a similar manner to the way temperature is defined in statistical mechanics, $T = \partial H / \partial S$. This simple approach makes it possible to calculate structural properties as expectation values by plugging the property into the partition function. For example, in consolidated materials made of grains, the volume associated with one grain is

$$\langle V_g \rangle = \frac{1}{Z} \int V_g(\{q\}) e^{-W(\{q\})/(\lambda x)} \Theta(\{q\}) \Pi dq \,. \qquad (2)$$

The Edwards conjecture has been investigated and experiments suggest that it works for granular materials [3,4]. While this theory has been proposed for rigid particle several works argue its suitability to foams [4,5]. If this is indeed the case this theory gives us, for the first time, a complete formalism that predicts properties of both disordered foams and granular materials.

In order to check and use the Edwards compactivity we must find first a set of parameters that describe correctly the system. The set of parameters must span the phase space of all possible configurations. It should have the correct number of parameters and should parameterize the volume function so W is written as a function of these properties.

In several recent publications [2,5] it was argued that a preferred set of such parameters is given by quadrons [5]. These are quadrilaterals in 2D and octahedra in 3D that are associated with cells and grains (or foam vertices). The construction of quadrons is based on the identification of the contacts between grains (or between vertices for foam) and identification of the cellular structure of the porous material. Here we propose to use the contact network to build a fully automatic method to obtain the cellular structure; to obtain the properties of the porous system; and to produce the quadrons. The coordination number, $Z_V$, of a basic element (vertex of foam or grain) is given by the number of contacts between the element and its nearest neighbors. For clarity and simplicity we will consider here the special case of $Z_V = 4$. Such systems include open and closed cell foams and general tetrahedral systems as silicates and dense colloidal aggregates.

## CHARACTERIZATION

The characterization is based on the partition of the void space into cells that are connected to each other via windows. The system is simplified into a polyhedral structure that captures the correct connectivity of both the lamella of the foam (or grains) and of the void space. To represent the system as such we need first to identify the plateau borders and vertices of the foam. There are various methods to do that [6,7] using geometrical methods, such as those that are based on the maximal ball algorithm [6,8] and topological methods as those based on the medial axis algorithm [7]. The maximal ball method assumes that pores are locally the widest parts of the system that are linked via narrow tube like throats. The algorithm and its successors are designed for low porosity and thus are ideal for the identification of the vertices and plateau borders. For foam the algorithm acts upon the lamella of the foam in which we the pores represent the vertices and throats represent the plateau borders. While this method is applicable to general structures, in what follows we concentrate on $Z_V = 3$ in 2D and $Z_V = 4$ foam in 3D. We also assume that the vertices and plateau borders have been properly identified.

Consider a $Z_v$=4 foam. The first stage of the characterization unifies the description of foam with the description of granular tetrahedral systems. Each vertex of the foam is transformed into a tetrahedron simply by connecting the middle of its four plateau borders to each other to form a pyramid (Figure 1). The tetrahedra touch their neighbors at the corners. Thus the corners are the contact points. Topologically, the foam and the dual tetrahedral-structure are homeomorphic. I.e., the connectivity of the two structures and of their void spaces are identical. In similar manner, the vertices of 2D foams are transformed into triangles by connecting the middle of the foam's ligaments.

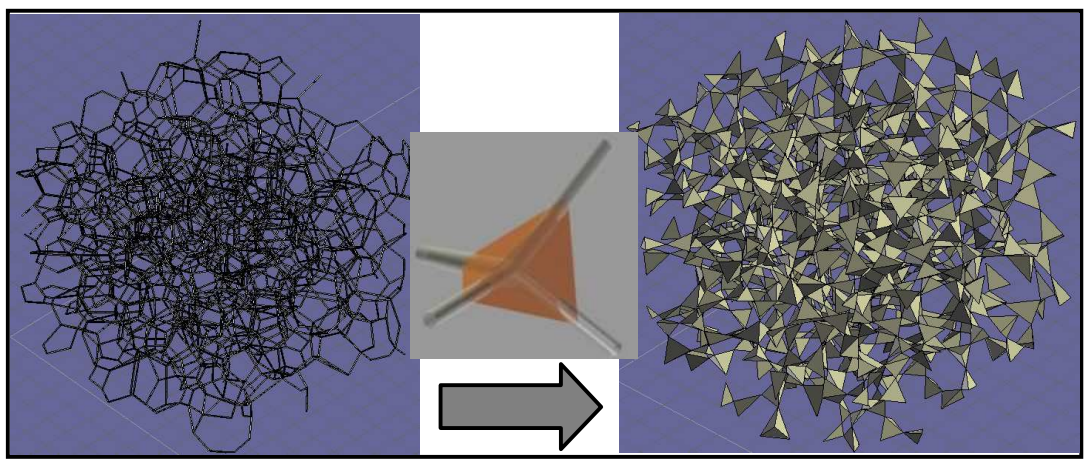

Figure 1. Left: The Plateau borders and vertices of 3D foam. Middle: Transformation of vertices to tetrahedra by connection of the middle of the plateau borders. Right: The dual tetrahedral network.

We continue with the definition of cells and windows. Cells are simply connected polyhedra (Fig 2). The cell boundary is composed of a set of facets of tetrahedra and of the windows associated with this set. The set of facets is closed. I.e. at all corners of each facet in the set there is a touching facet that belongs to the set. For a cell there can be only one facet per tetrahedron and each facet belongs to one cell. Thus there are exactly four cells around each of the tetrahedra. The cell defined by the faces is open and connected to its neighboring cells via the openings between the faces. Windows are these openings.

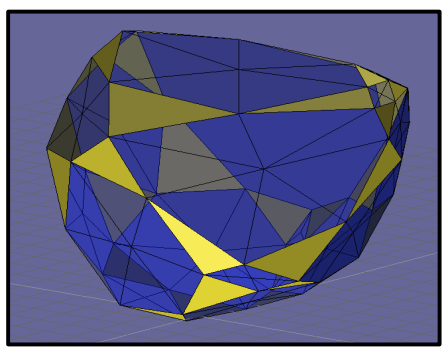

Figure 2. A typical cell. Its boundary is composed of a closed set of triangular facets of tetrahedra and windows that connect cells (transparent skewed polygons).

Windows are skewed polygons. Once the set of faces is known, the windows are identified by placing normal vectors, directed outwards, on each of the faces and ordering vectors around each face in the positive direction of rotation (Fig 3). Windows boundaries are identified by following the same vectors in the negative direction of rotation.

Thus we are left with the task of identifying the set of faces that belongs to a cell. In order to obtain the cellular structure we use a different set of algorithms from the one used for the lamella, because the maximal ball algorithm has been designed for low porosity. For foam the porosity is very high and the throats have degenerated from tube like structures to simply connected surfaces (i.e. windows). The algorithm used inflates a deformable balloon from the middle of a facet into the cell while keeping the balloon's curvature positive everywhere and at all times. This ensures that the expansion process stops once the balloon starts exiting the pore via the narrower throats (Figure 4).

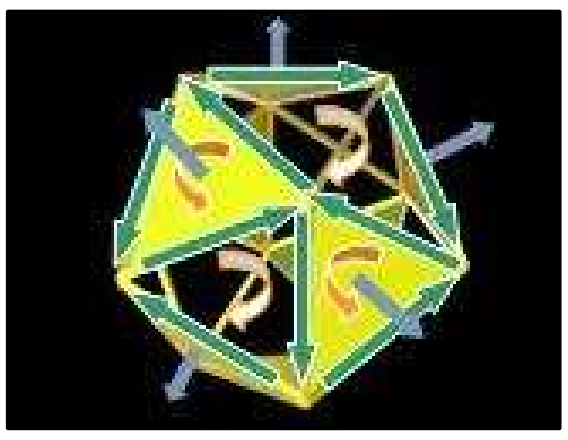

Figure 3. Extracting the throats: Normals placed on each face. Vectors ordered on the boundary of faces in the positive direction of rotation. The throat edge is extracted by following the same vectors in the negative direction of rotation.

The faces touching the balloon are added to a set of faces and the grown balloon serves as the basic search set for the facets that belong to the pore. Next the set of faces is checked for completeness. Faces having less than three neighbors are thrown recursively during the check until either a closed set is left or none are left. In the case that the set is not complete the neighboring faces to the ones in the list are added and completeness is checked again.

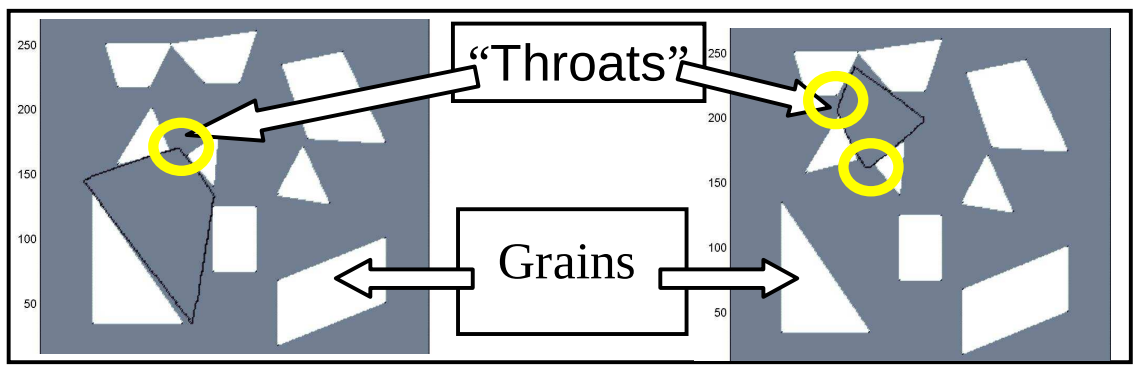

Figure 4. Extracting cells by expansion of Convex Deformable Balloon. As seen, the balloon (closed loop) is limited to the pore by convexity. Next the set of faces is identified and completed to form the cell.

As an example two samples were created, a 3D Voronoi tessellation made from Poisson distribution of seeding points [9] and foam – produced by the surface evolver [10]. For each, the tetrahedral structure has been produced and the cells identified. In figure 5 we depict the Voronoi system. Fig 5A depicts the tetrahedral structure. Figure 5B presents the cellular structure produced by the algorithm. Figure 5C shows the connectivity network of cells. In 5C spheres represent cells and tubes represent the windows.

Once the cellular structure is produce and connectivity obtained many properties of the cellular structure are extracted. The figure 6 depicts several properties of the Poisson-Voronoi tessellation and mono-dispersed foam.

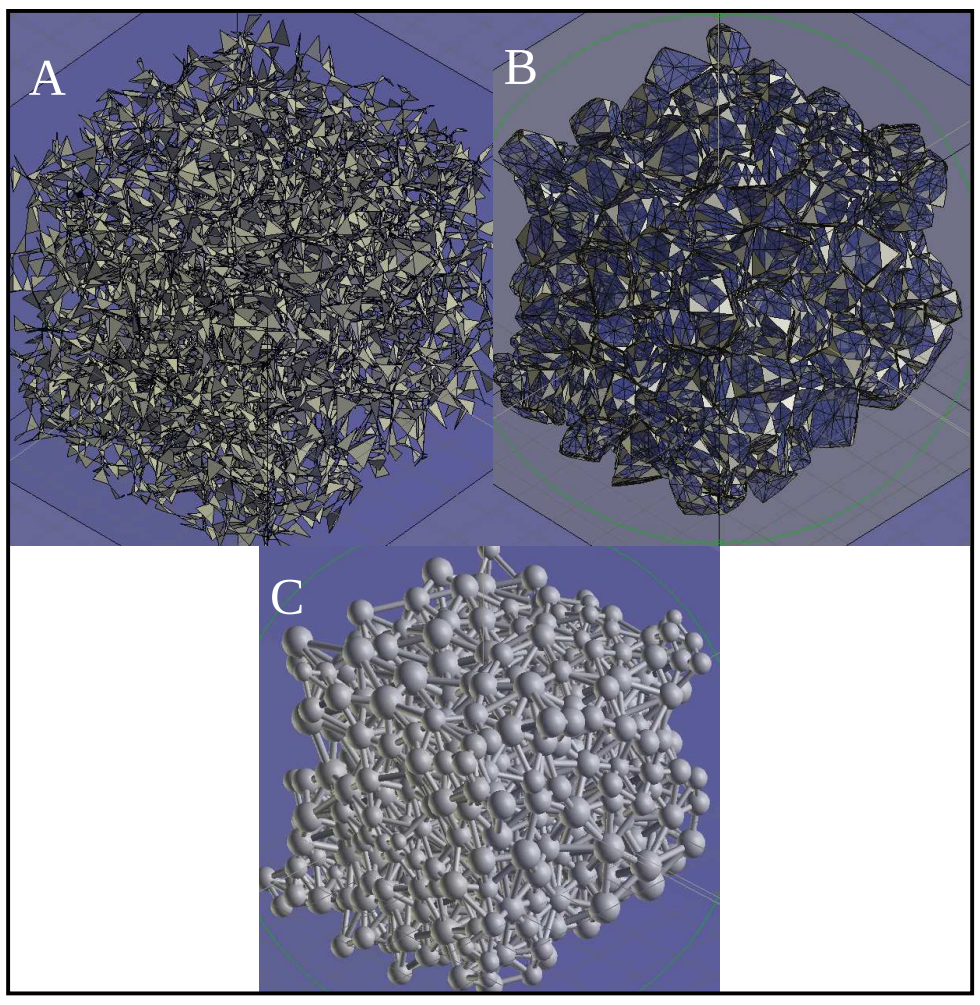

Figure 5. Voronoi example, extracting the cellular structure and connectivity of a tetrahedral structure.

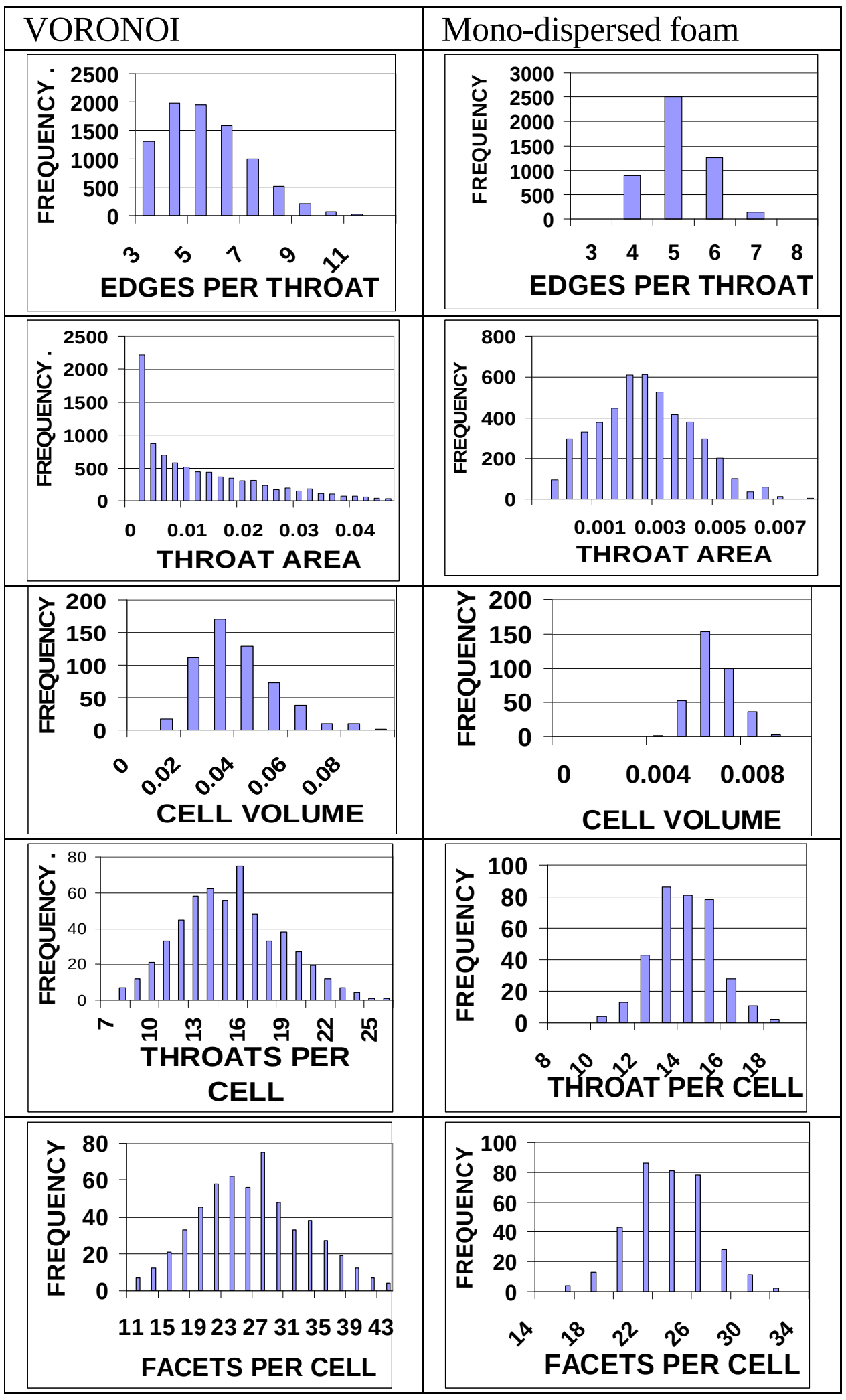

Figure 6: Extracting Data from the cellular structure. Right Voronoi tessellation, Left side: Mono-dispersed foam

While the topological properties of the foam structure, such as the number of throats per cell, can be extracted directly form the tetrahedral structure, the geometrical properties are only estimated. For example, knowing the tetrahedra that form a throat the original throat area can be extracted by introducing it to the maximal ball algorithm. The maximal ball algorithm extracts the plateau borders that correspond to these tetrahedra and the throat area can be extracted by the projection of the plateau borders onto the plain that is normal to the vector connecting the center of the two cells. In similar manners properties such as cell volume, surface area and the minimal cross section of contacts between grains can be extracted. However, we will leave those to a different publication due to the limited space here.

Last, the characterization above can be used to describe the basic structure found in the Edwards formalism. Due to lack of space this is done here only for a 2D system. For 2D foam vertices are replace by triangles (Fig. 7A). The parameters that describe the system are extracted for the division of the volume into quadrilaterals (Fig. 7B), termed quadrons [5]. For each edge between triangle t and cell c connect its edges to the centre of the triangle and centre of cell. These quadrilaterals tessellate space and thus $W = \sum V_{quadron}$, where $V_{quadron}$ is the volume of a quadron. Moreover, the number of quadrilaterals was shown to be equal to the number of degrees of freedom of the system. Thus, these volumes can serve as the degrees of freedom. Assuming that the quadrons are uncorrelated for simplicity the density of states is broken into multiplication of one quadron probability density function (PDF) $\Theta(\{V_q\}) = \prod P(V_{quadron})$ and the partition function is simplified to

$$Z = \left(\int e^{-V_q/\lambda X} P(V_q) dV_q\right)^N, \qquad (3)$$

where $N$ is the number of quadrons in the system. Extracting the PDF of quadron area for Voronoi tessellation we obtain a Gamma distribution: (Fig. 7C),

$$P(V_q) = b^a / \Gamma(a) \cdot V_q^{\,a-1} \exp(-bV_q).$$

As a simple example, the PDF is used in eq. (3) to obtain the n'th moment of the volume per quadron.

$$\langle V_q^{\,n} \rangle = \frac{\int e^{-V_q/X} V_q^{\,n} P(V_q) dV_q}{Z} = \frac{\Gamma(n+a)}{\Gamma(a)(b+1/X)^n}. \qquad (4)$$

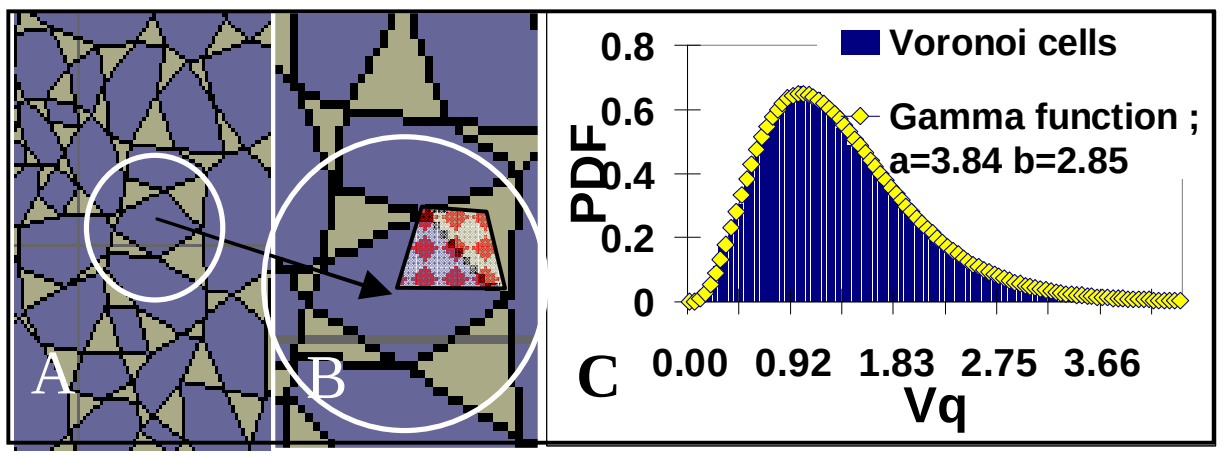

Figure 7: Left: part of a Voronoi structure. Vertices are transformed into triangles. Middle: quadron (quadrilateral) construction. Right: PDF of quadron volume.

## CONCLUSIONS

To conclude, we have demonstrated the applicability of an automated set of algorithms aimed at characterizing the cellular structure of foams and tetrahedral structures. The characterization is compatible with the Edwards entropic formalism and to demonstrate its use we considered a simplified uncorrelated system and calculated the expectation values of moments of the quadron areas. This is the first stage in the characterization. Taking into consideration the shape the plateau borders and vertices as additional degrees of freedom is the next stage.

## Acknowledgment

We wish to thank Stephen J Neethling for his help with foam production via surface evolver.